\documentclass[epj]{svjour}
\usepackage{graphicx}
\DeclareTextSymbol{\degre}{T1}{6}
\DeclareTextSymbol{\degre}{OT1}{23}
\begin{document}
\title{Magnetic relaxation measurements of exchange biased (Pt/Co) multilayers with perpendicular anisotropy}
\titlerunning{Magnetic relaxation of exchange biased (Pt/Co) multilayers}
\author{F. Romanens\inst{1} \and S. Pizzini\inst{1} \and J. Sort\inst{2} \and F. Garcia\inst{2} \and J. Camarero\inst{3} \and F. Yokaichiya\inst{1} \and Y. Pennec\inst{1} \and J. Vogel\inst{1} \and B. Dieny\inst{2}}
\authorrunning{F. Romanens}
\institute{Laboratoire Louis N\'{e}el \& IPMC, CNRS, 38340 Grenoble Cedex, France \and SPINTEC, URA CEA/DSM \& CNRS/SPM-STIC, CEA/Grenoble, 38054 Grenoble cedex 9, France \and Departamento de F\'{i}sica de la Materia Condensada, Universidad Aut\'{o}noma de Madrid, E-28049 Madrid, Spain}
\date{DOI: 10.1140/epjb/e2005-00053-3}
\abstract{Magnetic relaxation measurements were carried out by
magneto-optical Kerr effect on exchange biased (Pt/Co)$_{5}$/Pt/FeMn
multilayers with perpendicular anisotropy. In these films the
coercivity and the exchange bias field vary with Pt spacer
thickness, and have a maximum for 0.2 nm. Hysteresis loops do not
reveal important differences between the reversal for ascending and
descending fields. Relaxation measurements were fitted using
Fatuzzo's model, which assumes that reversal occurs by domain
nucleation and domain wall propagation. For 2 nm thick Pt spacer (no
exchange bias) the reversal is dominated by domain wall propagation
starting from a few nucleation centers. For 0.2 nm Pt spacer
(maximum exchange bias) the reversal is strongly dominated by
nucleation, and no differences between the behaviour of the
ascending and descending branches can be observed. For 0.4 nm Pt
spacer (weaker exchange bias) the nucleation density becomes less
important, and the measurements reveal a much stronger density of
nucleation centers in the descending branch. 
} 
\maketitle
\section{Introduction}
\label{intro}
Exchange-biased thin layer systems in which an antiferromagnet (AF) is in contact with a ferromagnet (FM) are still widely studied both for fundamental reasons and for their  applications in spin-electronics devices. In these systems a unidirectional anisotropy is induced by field cooling (FC) the bilayer system through the N\'{e}el  temperature of the AF. This leads to an increase of the coercivity of the FM layer and to a shift of the hysteresis loop, by the exchange bias field $E_{B}$, usually in the direction opposite to the cooling field. This effect is used in spin electronic devices to pin the magnetisation of the reference FM layer of a spin valve or a tunnel junction.  The complicated microscopic phenomena leading to exchange bias have been studied for more than 40 years, since the discovery of the effect by Meiklejohn and Bean \cite{Meikle56}.  A review of the main microscopic models proposed to explain exchange bias can be found in references \cite{Nogues1999,Berko1999,Stamps2000}. In order to correctly predict exchange bias fields and coercivity, the details of the FM/AF interface and the dynamics of the interface spins, frustrated by the competing interactions, should be taken into account. Models taking into account domain walls in the AF and in the FM \cite{Malozemoff,Mauri,Kiwi} and surface roughness and defects \cite{Takano} predict the right order of magnitude for $E_{B}$.

The presence of unidirectional interfacial FM/AF exchange interaction implies that the details of magnetisation processes in exchange biased films are different from those found for 'free' magnetic layers. Kuch et al. \cite{Kuch2002}  have clearly shown that the domain structure of cobalt grown on monocristalline FeMn films is strongly modified with respect to that of a free film. In particular the Co domains become much smaller and irregular. Quasi-static and dynamic magnetisation reversal in exchange biased  films have been studied by several groups. Dynamic coercivity, viscosity and spin dynamics  in exchange biased systems have been treated by Stamps \cite{StampsPRB2000,StampsJAP2001}. Thermal relaxation effects \cite{Taylor1999} and dynamic reversal \cite{Camarero1,Camarero2,Chopra2000} in Co/NiO films have been investigated by magneto-optical and transport methods. Thermally activated reversal in the AF layer has been widely studied by O'Grady group \cite{Hughes}. Different mechanisms of magnetisation reversal of the FM layer for fields applied against and with $E_{B}$ direction, showing up as asymmetric hysteresis loops, have been predicted theoretically \cite{Beckmann2003} and observed by several groups for exchange biased systems with in-plane magnetisation \cite{KirilyukJAP2002,McCordJAP2003,Gierlings2002,Fitzsimmons}. Gierlings \textit{et al.} \cite{Gierlings2002} deduced from polarised neutron reflectometry data on Co/CoO systems that magnetisation reversal in the descending branch (where the field is applied against the exchange bias direction) is due to nucleation and domain wall propagation, while the reversal in the ascending branch was interpreted as due to coherent rotation. Kerr microscopy data on CoFe/IrMn and Co/NiO \cite{KirilyukJAP2002,McCordJAP2003} showed that the magnetisation reversal involves nucleation and propagation in both branches. The volume of the sample reversing by nucleation or domain wall (DW) propagation is not the same for the two branches and the ratio between them depends on the angle between the applied field and the exchange bias direction.

In this paper we study the magnetisation reversal in exchange biased multilayers with perpendicular anisotropy in both ascending and descending fields. These systems are important from an application viewpoint as they are promising as ultra high density magnetic recording media \cite{Moritz2002,Landis2000} or as storage element in high density Magnetic Random Access Memory. Only a few investigations of perpendicular exchange biased systems, in general (Pt/Co) multilayers, are found in the literature \cite{Maat2001,Hellwig2002,Kagerer2000,Liu2003,Zhou2004,Garcia2002,Garcia2003a,Garcia2003b,Sort2003,Sort2004}. Unbiased M/Co/M trilayers and multilayers with M~=~Pt,~Pd and Au  have been studied to clarify the origin of perpendicular anisotropy and its relation to enhanced interfacial orbital moments and anisotropies \cite{Bruno,Weller}. Magnetisation dynamics in Pt/Co/Pt and Au/Co/Au trilayers has been widely investigated by Ferr\'{e} \textit{et al.} by Kerr microscopy \cite{Ferre2002}. The variation of the domain structure with the amplitude of the applied field has been recently studied by  Woodward \textit{et al.} \cite{Woodward2003}.

In a previous paper \cite{Garcia2002} we have reported the study of the dynamic coercivity  of a serie of (Pt/Co)$_ 4$/FeMn multilayers. Magneto-optical Kerr measurements suggested that for low sweep rates of the applied field the reversal of the (Pt/Co)$_4$ magnetisation was initiated by the nucleation of reversed domains but dominated by propagation of domain walls. Although other measurements on (Pt/Co) multilayers did not show a difference in the reversal mechanism in the ascending and descending branches of the hysteresis loops \cite{Hellwig2002} our fits of the $H_{C}$ \textit{vs.} $dH/dt$ curves revealed a difference in the reversal mode of the two branches. The Barkhausen volume associated to the reversal in the descending branch (against the exchange bias direction) was found to be smaller than the one for the ascending branch. This suggested that the density of pinning centers hindering the domain wall motion was larger for the descending branch than for the ascending branch.

In this paper we present magnetisation relaxation measurements on  exchange biased (Pt/Co)$_5$/Pt/FeMn multilayers, in which a thin Pt spacer is grown between the last Co and the FeMn layers. The effect of the Pt spacer on the magnetic properties of the multilayer has been reported by Garcia \textit{et al.} \cite{Garcia2003b}. The presence of a thin Pt spacer in contact with FeMn increases the effective perpendicular anisotropy of the topmost Co layer and leads to an increase of the exchange bias field and coercivity.

Our measurements reveal that the mechanisms leading to magnetisation reversal strongly depend on the thickness of the Pt spacer. As already observed for Pt/Co/Pt trilayers \cite{Ferre2002}, domain wall propagation dominates the magnetisation reversal of unbiased samples which in our case correspond to samples with thick Pt spacers. For thin Pt spacers, and in the presence of exchange bias,  the reversal is instead dominated by domain nucleation and an asymmetry is observed between the reversal in the two branches of the hysteresis loop for a sample with moderate exchange bias.

\section{ Experimental methods and analysis }
\label{sec:exp}

We carried out measurements on  (Pt(2nm)/Co(0.4nm))$_5$/ Pt(\textit{t})/FeMn(13nm) multilayers with \textit{t}= 0, 0.2, 0.4, 0.7 and 2 nm. Multilayer samples were deposited on thermally oxidised Si wafers by dc magnetron sputtering and capped  with 2~nm of Pt. They are polycrystalline with a weak (111) texture. The samples were field cooled from 150\degre{}C, above the blocking temperature, under a field of 0.25 T applied perpendicular to the film plane. The details of  sample preparation and magnetic properties are presented in references \cite{Garcia2002,Garcia2003b}. Samples show perpendicular exchange bias field $E_{B}$, enhanced coercivity $H_{C}$,  and a perpendicular magnetic anisotropy which depend on the Pt spacer thickness (Figure \ref{Fig:bias-et-anisotropie}). The anisotropy energies measured with a vibrating sample magnetometer (VSM)  were deduced from the value of  the saturation magnetisation obtained by applying a magnetic field in the plane of the layers.

Hysteresis loops of the multilayer samples were  measured  at room temperature using polar Kerr effect.  Relaxation measurements were carried out in the following way. The sample magnetisation was first saturated to $+M_{S}$ by applying a strong field out of the plane of the sample. At a time $t$=0 an opposite field $H<H_{C}$,  close to the quasi-static coercivity,  was applied. While the magnetisation relaxes from $+M_{S}$ to $-M_{S}$, the temporal variation $M(t)$ of the magnetisation with constant applied field is measured by Kerr effect. This is done for the two branches of the hysteresis cycle, and for several applied field amplitudes giving relaxation times between some milliseconds and some seconds.

The shapes of the $M(t)$ curves can be understood in the light of the theory developed by Fatuzzo \cite{Fatuzzo1962} which was applied for the first time to magnetic materials by Labrune \textit{et al.} \cite{Labrune1989}. The reversal is assumed to be thermally activated and to occur first by random nucleation, according to a statistical process having probability $R$ per unit time~:
\begin{equation}
\label{EqN}
N=N_{0}(1 - \exp (- Rt ))
\end{equation}
where $N$ is the total number of nuclei at time $t$ and  $N_{0}$ the total number of nucleation sites.

Each circular domain of initial radius $r_c$ is assumed to grow with a constant radial velocity $v$. It can be shown that the fractional area $B(t)$ whose magnetisation has not reversed at time $t$ is given by~:

\begin{eqnarray}
\label{EqFatuzzoModel}
 B(t) = & \exp \Big{[} -2k^2 \Big{(}  1- (Rt+k^{-1}) + \frac{1}{2} (Rt + k^{-1})^2 \nonumber \\
 & - \exp (-Rt) (1-k^{-1}) - \frac{1}{2} k^{-2} (1-Rt) \Big{)} \Big{]}
\end{eqnarray}

where $k=v/Rr_c$. The shape of the relaxation curve then depends only on the parameter $k$ which expresses the relative weight of the sample volume reversed by nucleation and propagation. When domain wall propagation dominates (for $k \gg 1$), equation \ref{EqFatuzzoModel} can be reduced to~:
\begin{equation}
      B(t) \simeq \exp(- k^2 R^3 t^3 / 3)
\end{equation}
and it gives rise to a S-shaped variation of the magnetisation. When nucleation governs magnetisation reversal, for small values of $k$ ($k \ll 1$), then
\begin{equation}
      B(t) \simeq \exp(-R t)
\end{equation}
and the magnetisation has an exponential decay.

Note that one of the limitations of Fatuzzo's model is that it considers the existence of only one energy barrier for each reversal process: $E_p$ and $E_n$ associated with domain wall propagation and nucleation mechanisms.

\section{Results and discussion }
\label{sec:res-disc}

The hysteresis loops measured for the multilayer samples show that: i) the exchange bias field and the coercivity strongly depend on Pt spacer thickness and ii) no clear differences are present between the shapes of the descending and ascending branches.

\begin{figure}
\includegraphics[width=\columnwidth]{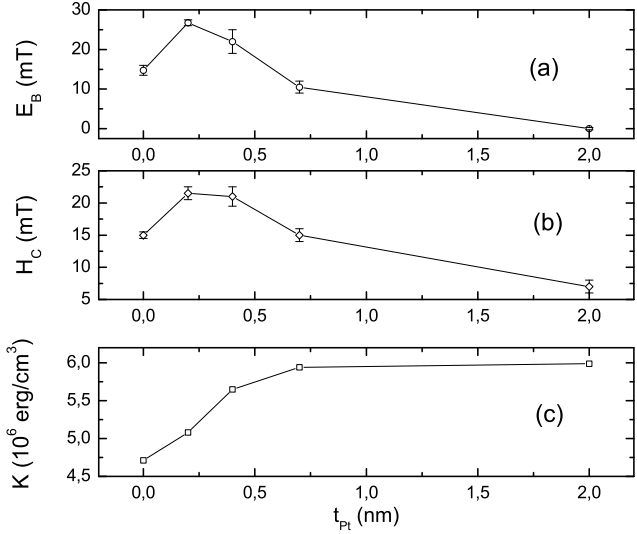}
\caption{Values of exchange bias field (a), coercivity (b) and anisotropy energy (c) as a function of Pt spacer thickness, for the (Pt/Co)$_{5}$/Pt$(t)$/FeMn multilayers. The line is a guide for the eyes.}
\label{Fig:bias-et-anisotropie}
\end{figure}

As shown in figure~\ref{Fig:bias-et-anisotropie} the coercivity and the exchange bias field increase for thin Pt spacer, pass through a maximum for about 0.2~nm of Pt and decrease for thicker  Pt spacer where $E_{B}$ becomes zero. Garcia et al.  \cite{Garcia2003b} attributed the increase of $E_{B}$ and $H_{C}$, obtained with thin Pt spacers, to the increase of the perpendicular component of the magnetisation of the topmost Co layer and, through the exchange interaction between different Co layers, to the increase of the perpendicular anisotropy of the whole multilayer. The perpendicular anisotropy strongly increases for small Pt spacer thickness, then saturates for about 0.7~nm Pt. The increase of the perpendicular anisotropy is also reflected by the shape of the hysteresis curves. For thin Pt spacers the cycles are tilted -- signature of an in-plane magnetisation component -- while for 0.7~nm  and 2~nm of Pt the curves are practically square.

The increase of the coercivity $H_{C}$ and of the exchange bias field $E_{B}$ \cite{Garcia2003b} can then be explained. When the Co moments are better aligned along the out-of-plane direction, their projection along the AF moments direction increases, since at their turn these moments are on average aligned along the easy axis closest to the (perpendicular) FC direction. The presence of a maximum in $E_{B}$ and $H_{C}$ followed by an exponential decrease is due to the competition between two effects: i)~the anisotropy of the Co layers which increases with Pt thickness and ii)~the short range FM-AF interaction that, for a non magnetic spacer, induces an exponential decay of coercivity and exchange bias \cite{Garcia2003b}.

\begin{figure}
\includegraphics[width=\columnwidth]{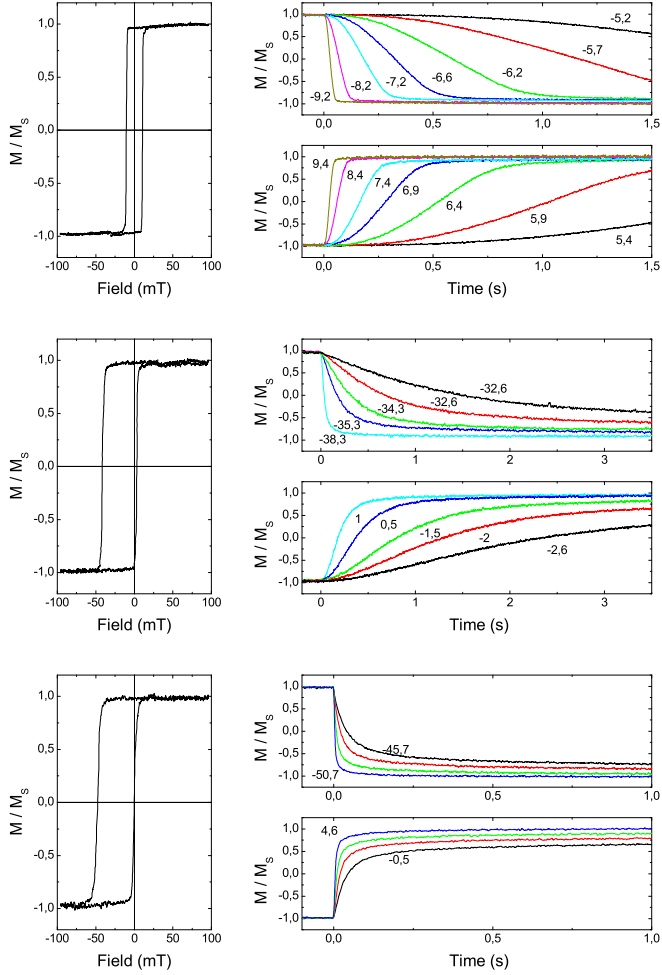}
\caption{Normalised hysteresis cycles (left curves) and associated magnetisation relaxation measurements $M(t)$ (right curves) measured for the three (Pt/Co)$_{5}$/Pt$(t)$/FeMn with Pt spacer thickness $t$ = 2~nm Pt (sample I - top), 0.4~nm Pt (sample II - middle) and 0.2~nm Pt (sample III - bottom). The hysteresis loops were measured with a field sweep rate of 50~mT/s. The relaxation curves were measured starting from positive (negative) saturation and for several opposite fields close to the coercive field of the descending (ascending) branches. These are shown respectively in the top (bottom) panel beside the corresponding hysteresis loops.}
\label{Fig:cycles-et-relax}
\end{figure}

The relaxation measurements for the three multilayer samples with Pt spacer thickness $t$ = 2, 0.4 and 0.2 nm, called from now on samples I, II and III, are shown in figure~\ref{Fig:cycles-et-relax} together with the corresponding hysteresis loops. After saturation of the sample in the positive (negative) direction, a magnetic field $H$ of opposite direction and amplitude close to the coercive field in the descending (ascending) branches is applied at time $t$=0. The time dependence of the magnetisation is measured for several values of $H$.  The $M(t)$ curves have different shapes for the three samples, revealing that the reversal occurs by different mechanisms depending on the Pt layer thickness. Qualitatively, according to Fatuzzo's model, the S-shaped curve obtained for sample I ($t=2$~nm and $E_{B} =0$) indicates that the reversal is dominated by domain wall propagation. On the other hand, for samples II and III,  which exhibit exchange bias, the $M(t)$ curves have a much more exponential decay, revealing that nucleation dominates the reversal. Note also that a difference between the shape of the relaxation curves measured for applied fields in the descending and ascending branches is clearly visible for sample II.

\begin{figure}
\includegraphics[width=\columnwidth]{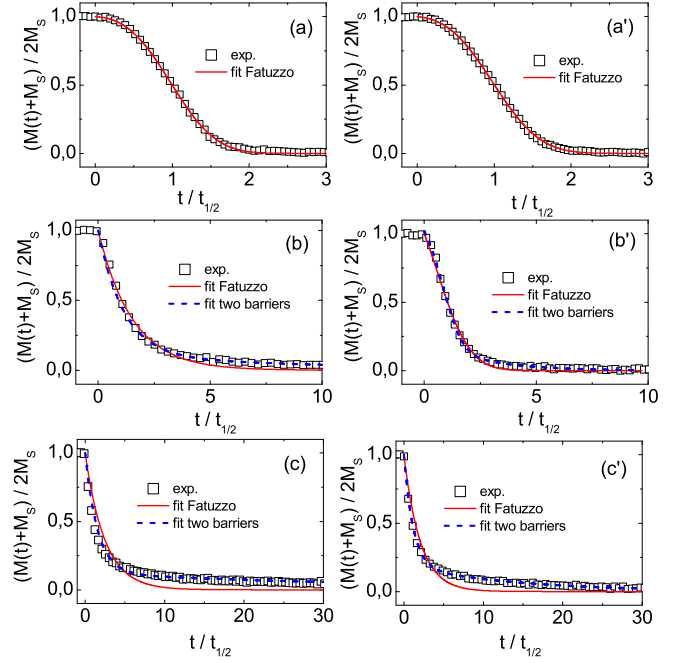}
\caption{Relaxation measurements (symbols) measured for samples I (top), II (middle) and III (bottom) for the descending (a,b,c) and ascending (a',b',c') branches of the hysteresis loops. Full lines represent the fits using Fatuzzo's model (\ref{EqFatuzzoModel}). The dotted lines are the fits considering the presence of two energy barriers.}
\label{Fig:fit-relax}
\end{figure}

For sample I ($E_{B}=0$) the fit of the relaxation curve to the expression (\ref{EqFatuzzoModel}) gives $k$ values around 100 (figure \ref{Fig:fit-relax}).  This confirms that magnetisation reversal proceeds essentially by propagation of domain walls starting from a few reversed domains. The quality of the fit suggests that only one energy barrier (or a narrow distribution of energy barriers) is associated to the thermally activated process. The relaxation curves measured for the various magnetic field amplitudes and plotted as a function of the reduced time $t/t_{1/2}$ (where $t_{1/2}$  is the time for which the magnetisation decreases to 50\%  of its saturation value) can all be superposed. This shows that for this sample and within the range of fields used here, the $k$ value deduced from Fatuzzo's model, and therefore the ratio between domain wall propagation and nucleation reversal mechanisms, is sensitively independent of the field. This also means that $v$ and $R$ show the same exponential variation with applied field.

\begin{figure}
\includegraphics[width=\columnwidth]{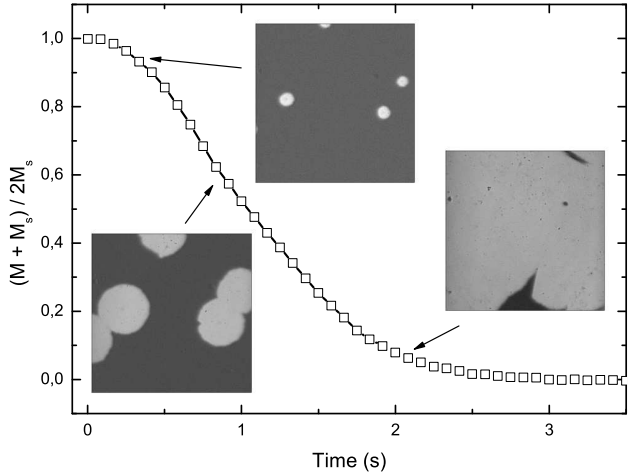}
\caption{Relaxation curve and corresponding domain structure measured for a (Pt/Co)$_4$ multilayer without antiferromagnet. The S-shaped relaxation curve is typical of reversal dominated by domain wall propagation. The Kerr microscopy images confirm that the reversal proceeds by nucleation of a few domains, followed by propagation of their domain walls.}
\label{Fig:relax-et-image}
\end{figure}

To confirm the validity of our interpretation of the relaxation curves within Fatuzzo's model, in figure \ref{Fig:relax-et-image} we present the results of time-dependent Kerr microscopy measurements obtained for a sample similar to sample I, a (Pt/Co)$_4$ multilayer grown without FeMn overlayer. The magnetisation reversal in the two samples is expected to be similar, since we have seen that the 2 nm thick Pt spacer in sample I cancels the effects of the AF/FM interaction.  As for the macroscopic relaxation measurements, the sample was saturated in the positive direction, and at time $t=0$ a negative magnetic field was applied. Kerr images were then acquired for several times between $t=0$ and the time where negative saturation is obtained.  For this sample the fit of the relaxation curve with Fatuzzo's model gave $k=10$ which indicates that propagation of domain walls dominates the reversal.  Kerr images confirm that only a few domains nucleate within the microscope field of view (250 $\mu$m $\times$ 250 $\mu$m) and propagation of domain walls leads to negative saturation. Images of the magnetic domain structure of sample I could not be obtained with Kerr microscopy because of the 13 nm thick FeMn overlayer which strongly reduced the signal. Since a value of $k=100$ was found from relaxation measurements, we expect for sample I even less nucleation sites than for the (Pt/Co)$_4$ multilayer.

For the exchange biased samples II and III, the relaxation curves could not be fitted over the whole timescale using Fatuzzo's expression (i.e. considering a single energy barrier for the thermally activated process) since for long relaxation times  the magnetisation varies more slowly than predicted within this model (see figure 3).

The experimental curves could only be fitted correctly when at least two energy barriers were considered, one corresponding to the fast reversal described by Fatuzzo's model and one associated with the slower reversal of the last 20-30\% of the samples magnetisation.

The results of the fits using one or two energy barriers are presented in figure 3. For sample III  the fast decay of the magnetisation, for both hysteresis branches, could be fitted using an exponential $B(t)$ curve with $k=0$. This confirms that in this sample, presenting maximum exchange bias, the initial stages of the reversal are dominated by nucleation of reversed domains. Note that for this sample no information about a possible difference in the reversal in the two hysteresis branches can be obtained - for small $k$ values (large nucleation density) the shape of the relaxation curves depends very little on $k$, and differences in $k$ values for the two branches cannot be reliably extracted from the fits.

On the contrary, for sample II the initial stages of the relaxation are very different for the two branches of the hysteresis loops. In the ``descending'' branch (reversal against $E_{B}$) the relaxation is dominated by nucleation ($k \simeq 0$) while for the ``ascending'' branch (reversal toward the direction of $E_{B}$) propagation is more favorable and the relaxation proceeds by the formation of a smaller number of nucleation centers ($k \simeq 5$). The results for this sample indicate that the reduction of the exchange bias leads to a reduction of the density of nucleation centers (larger $k$ values). The differences in the reversal mechanisms in the two branches then become clearly observable, since the $M(t)$ curves change more for larger k values. In both samples II and III the reversal of the last 20-30\% of the total magnetisation proceeds slower than in Fatuzzo's model, with a time constant which  depends on the amplitude of the applied field and which is much larger than that of the initial decay curve. This suggest that this ``hard volume'' reverses by overcoming an energy barrier which is much larger than that associated with the initial thermally activated process. The fractional volume reversing with a long time constant is more important for sample III (larger exchange bias) than for sample II, and for both samples is larger for the descending branch, where reversal proceeds via the formation of more nucleation centers, than for the ascending branch. This asymmetry in the relaxation curve can be seen for both samples.

Similar effects in the relaxation curves have been found by Pommier \textit{et al}. \cite{Pommier1990} for Au/Co/Au trilayer samples, for which the reversal proceeded mainly by domain wall propagation. In that case the slow relaxation of the magnetisation for long timescales was attributed to the presence of hard magnetic centers which pinned the domain wall motion therefore preventing a rapid reversal. The presence of hard magnetic centers may not be excluded in our samples. However the correlation between the 'hard' fractional volume and the amplitude of the exchange  bias field is rather striking and suggests a different explanation. As we said already, the fractional volume left unreversed increases with the exchange bias (it is zero for sample I  where propagation dominates and is maximum for sample III where nucleation density is maximum)  and is larger for the descending branch, where the nucleation dominates. This suggests that the slow decay of the relaxation curve in the late stages of reversal may be related to the formation of a large density of 360\degre domain walls which may require high fields to be evacuated from the sample. Since the volume associated to these walls increases with the nucleation density, the relationship between  unreversed volume and exchange bias may be explained.

Changes in magnetisation reversal mechanisms of exchange biased FM layers with respect to free layers are induced by the interaction at the FM/AF interface. This interaction has both structural and magnetic origins. Structural changes, like the increase of the density of defects,  may explain the increase of the nucleation density (and the barrier for DW motion) observed in this study in samples II and III with respect to sample I.

To explain the difference in reversal mechanism for the two hysteresis branches, however, the magnetic FM/AF interaction has to be taken into account. In this sense, the main result of this work is the observation of a larger number a nucleation centers in the descending branch of the hysteresis cycle, where the field is applied opposite to the exchange bias field.

According to Nikitenko \textit{et al.} ~\cite{Nikitenko2000} the different reversal for the two hysteresis branches can not be explained for the case of an uncompensated AF spin structure, since in this case the pinning centers for DW motion are the same for increasing and decreasing fields. The asymmetry is a manifestation of the inhomogeneity of the local uniaxial anisotropy, already postulated by Malozemoff ~\cite{Malozemoff}. For field applied against the exchange bias direction the magnetic interface interaction acts, on average, against the reversal leading to a higher energy barrier for DW motion and therefore favouring reversal by nucleation. Differences in the initial nucleation process in the two hysteresis branches of (Co/Pt)$_{50}$ multilayers on CoO, associated to the presence of local regions more exchange biased than the average, had been observed by Hellwig \textit{et al.} \cite{Hellwig2002}. No differences between the reversal for the two branches were however observed, since the latter was determined by the strong dipolar fields characteristic of this thick multilayer system.

It could be argued that the different nucleation density observed for the different samples could be simply related to their different perpendicular anisotropy. However, measurements on multilayer samples presenting a thinner AF overlayer (therefore no exchange bias) have shown that changes in the nucleation density related to changes in anisotropy are at least one order of magnitude smaller than those observed for exchange biased samples.

To conclude, we have shown for thin multilayer samples presenting perpendicular anisotropy, that the magnetisation reversal is strongly modified in exchange-biased films (thin Pt spacers) with respect to the unbiased film (large Pt spacer). Differences in the reversal of the magnetisation in the descending and ascending branches of the hysteresis loops can also  be observed. These differences are not observed in the hysteresis cycles, but become evident in relaxation curves, as soon as the nucleation rates are not too high. Similar effects had been observed by dynamic coercivity data in (Pt/Co)$_4$/FeMn samples ~\cite{Garcia2002} without Pt spacer and can be attributed to a distribution of the exchange bias strength. One of the future objectives of this work is to clarify  how differences in the reversal in the descending and ascending branches are related to the average strength of the perpendicular anisotropy and to the strength of the exchange bias field.

\section*{Acknowledgement}
This work was partially supported by the European Community through the RTN grant NEXBIAS.

\end{document}